\documentclass[graybox]{svmult}

\usepackage{mathptmx}       % selects Times Roman as basic font
\usepackage{helvet}         % selects Helvetica as sans-serif font
\usepackage{courier}        % selects Courier as typewriter font
\usepackage{type1cm}        % activate if the above 3 fonts are
\usepackage{makeidx}         % allows index generation
\usepackage{graphicx}        % standard LaTeX graphics tool
\usepackage{multicol}        % used for the two-column index
\usepackage[bottom]{footmisc}% places footnotes at page bottom

\makeindex             % used for the subject index
\begin{document}

\title*{Cosmic-ray-induced ionization in molecular clouds adjacent to supernova remnants}
\author{F.~Schuppan, J.~Becker~Tjus, J.H.~Black, S.~Casanova and M.~Mandelartz}
\institute{F.~Schuppan \at Ruhr-Universit\"at Bochum, Fakult\"at f\"ur Physik \& Astronomie, 44780 Bochum, Germany \\ 
\email{florian@tp4.rub.de}
\and J.~Becker~Tjus \at Ruhr-Universit\"at Bochum, Fakult\"at f\"ur Physik \& Astronomie, 44780 Bochum, Germany
\and J.H.~Black \at Dept.\ of Earth and Space Sciences, Chalmers University of Technology, Onsala Space
Observatory, SE-439 92 Onsala, Sweden
\and S.~Casanova \at Unit for Space Physics, North-West University, Potchefstroom 2520, South Africa
\and M.~Mandelartz \at Ruhr-Universit\"at Bochum, Fakult\"at f\"ur Physik \& Astronomie, 44780 Bochum, Germany}
\maketitle

\abstract*{Energetic gamma rays (GeV to TeV photon energy) have been
    detected toward several supernova remnants (SNR) that are associated
		with molecular clouds. If the gamma rays are
		produced mainly by hadronic processes rather than leptonic processes
		like bremsstrahlung, then the flux of energetic cosmic ray nuclei ($>1$~GeV)
		required to produce the gamma rays can be inferred at the site
		where the particles are accelerated in SNR shocks. It is of great
		interest to understand the acceleration of the cosmic rays
		of lower energy ($<1$~GeV) that accompany the energetic component.
		These particles of lower energy are most effective in ionizing
		interstellar gas, which leaves an observable imprint on the
		interstellar ion chemistry. A correlation of energetic gamma
		radiation with enhanced interstellar ionization can thus be used to
		support the hadronic origin of the gamma rays and to constrain the
		acceleration of ionizing cosmic rays in SNR.}
		
\abstract{Energetic gamma rays (GeV to TeV photon energy) have been
    detected toward several supernova remnants (SNR) that are associated
		with molecular clouds. If the gamma rays are
		produced mainly by hadronic processes rather than leptonic processes
		like bremsstrahlung, then the flux of energetic cosmic ray nuclei ($>1$~GeV)
		required to produce the gamma rays can be inferred at the site
		where the particles are accelerated in SNR shocks. It is of great
		interest to understand the acceleration of the cosmic rays
		of lower energy ($<1$~GeV) that accompany the energetic component.
		These particles of lower energy are most effective in ionizing
		interstellar gas, which leaves an observable imprint on the
		interstellar ion chemistry. A correlation of energetic gamma
		radiation with enhanced interstellar ionization can thus be used to
		support the hadronic origin of the gamma rays and to constrain the
		acceleration of ionizing cosmic rays in SNR. Using observational 
		gamma ray data, the primary cosmic ray proton spectrum can be 
		modeled for $E$~$>1$~GeV, and careful extrapolation of the spectrum 
		to lower energies offers a method to calculate the ionization rate 
		of the molecular cloud.}

%\abstract{Each chapter should be preceded by an abstract (10--15 lines long) that summarizes the content. The abstract will appear \textit{online} at \url{www.SpringerLink.com} and be available with unrestricted access. This allows unregistered users to read the abstract as a teaser for the complete chapter. As a general rule the abstracts will not appear in the printed version of your book unless it is the style of your particular book or that of the series to which your book belongs.\newline\indent
%Please use the 'starred' version of the new Springer \texttt{abstract} command for typesetting the text of the online abstracts (cf. source file of this chapter template \texttt{abstract}) and include them with the source files of your manuscript. Use the plain \texttt{abstract} command if the abstract is also to appear in the printed version of the book.}

\section{Introduction}
\label{sec:1}
\begin{figure}
	\centering
		\includegraphics[scale=1.0]{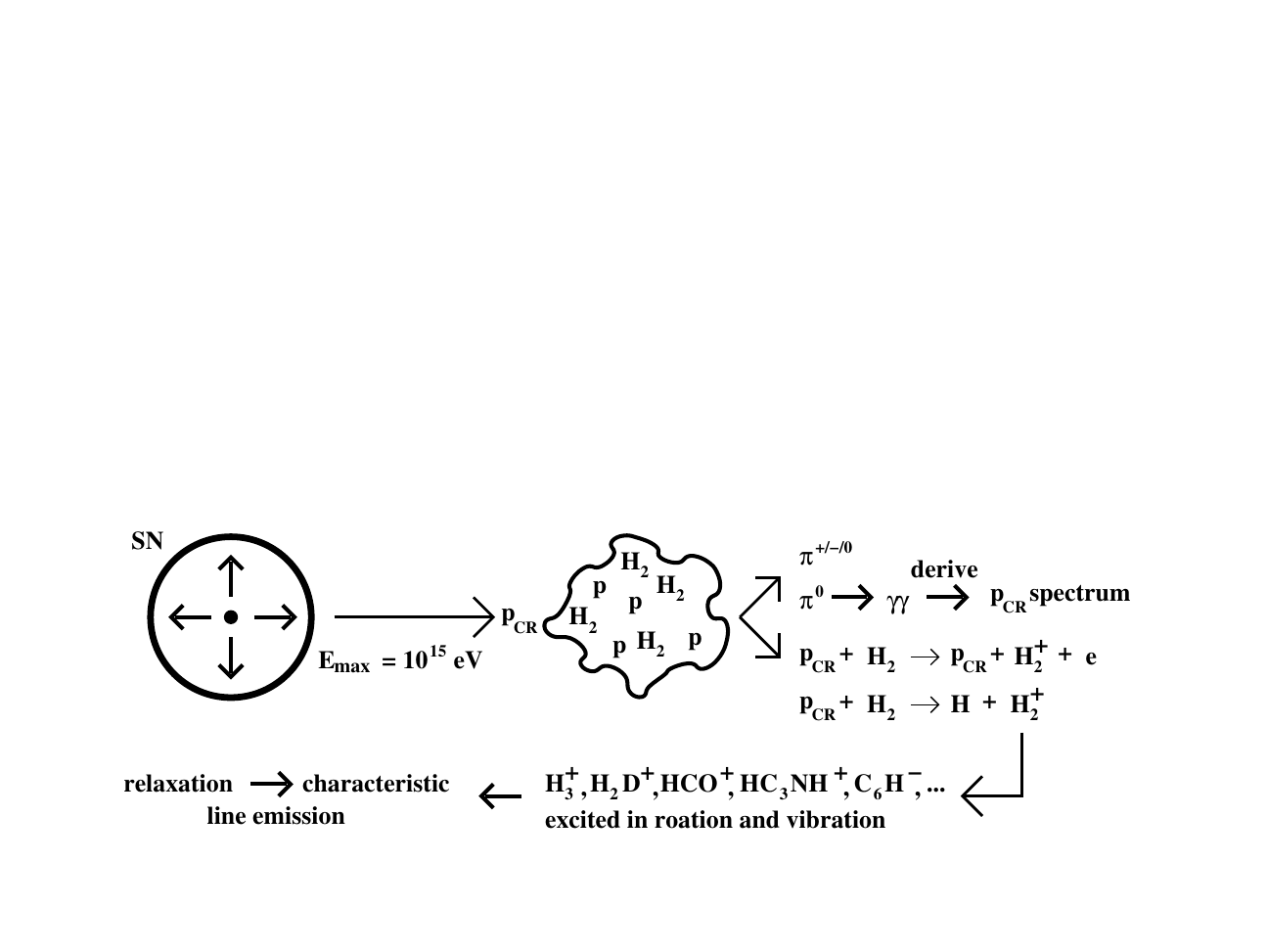}
	\caption{Sketch of the basic idea}
	\label{scheme}
\end{figure}

The origin of cosmic rays (CRs) is one of the central questions in modern astrophysics. In particular, for gamma rays the formation processes are still a matter of debate due to ambiguity of the observations. The detection of the emission of gamma rays in GeV and TeV towards molecular clouds in the direct vicinity of supernova remnants (e.g.\ \cite{abdo(W51C)2009}) has further fueled the question about the origin of this radiation. In principle, there are three different scenarios by which these particular gamma rays can be formed: bremsstrahlung or inverse Compton scattering of electrons in a leptonic scenario or the formation and -- in turn -- decay of neutral pions from scattering of CR protons on ambient protons in a hadronic scenario. In most cases, it is not possible to determine which process actually is at work based on the gamma ray spectrum alone. Therefore, correlation studies need to be performed. 
In the hadronic scenario, CR protons of energies $E~>~$1~GeV form neutral pions by deep inelastic scattering on ambient matter, which in turn decay into two gamma rays. It is to be expected that there are also lower energy protons, which are most effective in ionizing hydrogen. Using molecular clouds as tracers for these ionizing lower energy CR protons in spatial correlation with GeV to TeV gamma radiation could offer a strong hint at hadronic origin of the gamma rays, since CR electrons do not penetrate molecular clouds as deeply as CR protons, resulting in a different ionization profile. The ionized molecular hydrogen, H$_2^+$, triggers a chemical network, forming about 15 to 20 molecular ions, which are formed in ro-vibrationally excited states (see e.g.\ \cite{mccarthy2006, black2007}). If there is a sufficient amount of these ions formed, i.e.\ the ionization rate is high enough, the relaxation of these ions results in characteristic line emission, which can be detected as tracer for enhanced ionization \cite{becker2011}. 
The basic idea is sketched in figure \ref{scheme}.

%Use the template \emph{chapter.tex} together with the Springer document class SVMono (monograph-type books) or SVMult (edited books) to style the various elements of your chapter content in the Springer layout.

%Instead of simply listing headings of different levels we recommend to
%let every heading be followed by at least a short passage of text.
%Further on please use the \LaTeX\ automatism for all your
%cross-references and citations. And please note that the first line of
%text that follows a heading is not indented, whereas the first lines of
%all subsequent paragraphs are.

\section{Deriving the CR proton spectrum}
\label{sec:2}
\begin{figure}
	\centering
		\includegraphics[scale=0.7]{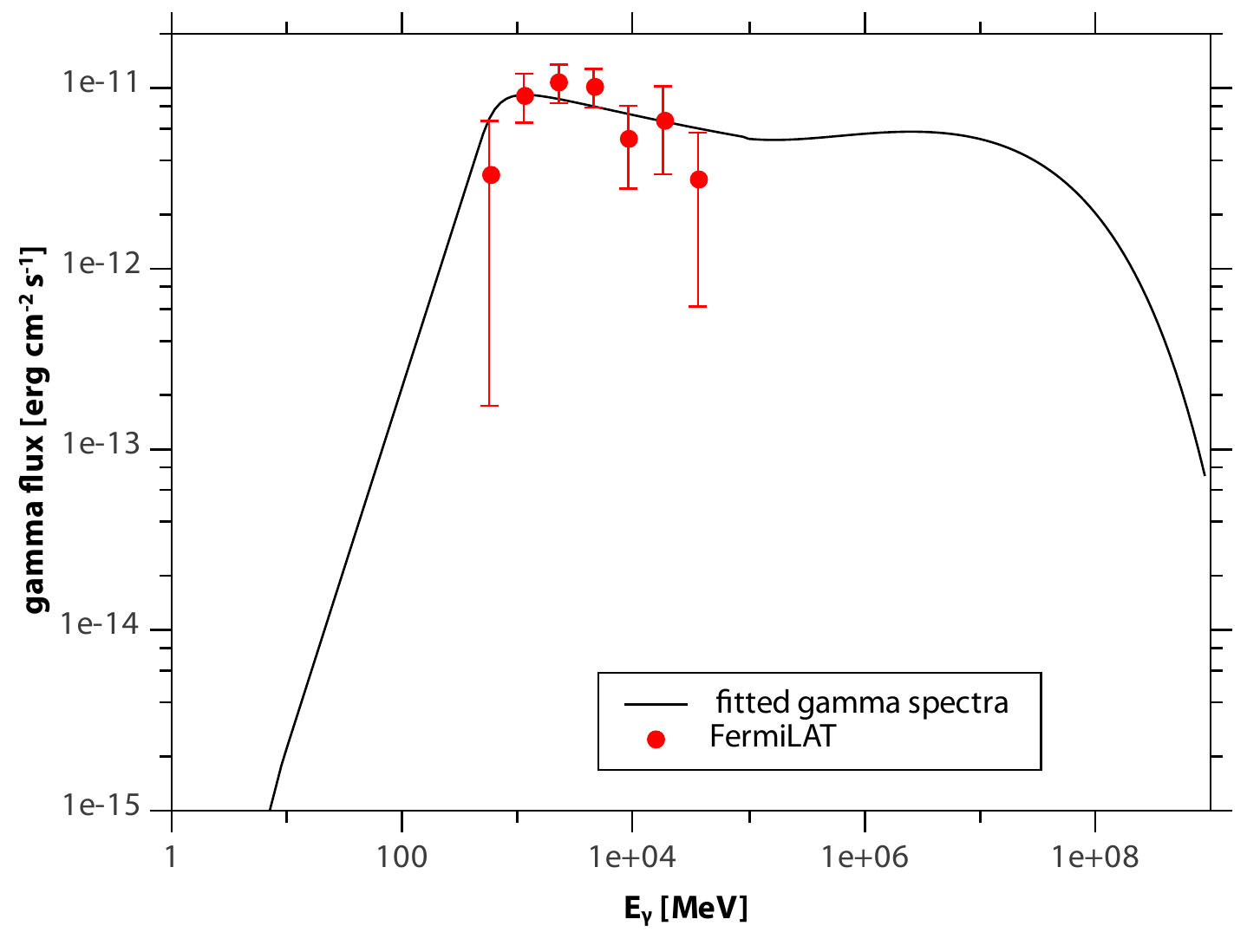}
		\caption{Gamma ray spectrum from pion decay fitted to gamma ray measurements for G349.7+0.2. Gamma ray measurements from \cite{castro2010}. Primary CR proton spectral shape assumed: $j_{\rm p}(E_{\rm p}) = a_{\rm p}\left(\frac{E_{\rm p}}{\mathrm{1~TeV}}\right)^{-\alpha_l}\exp{\left(-\frac{E_{\rm p}}{\mathrm{1~PeV}}\right)}$. Fit parameters: $\chi^2/{\mathrm{dof}}~=~0.88638$, $a_{\gamma}~=~(820.97822~\pm~494.07024)/\left(\frac{n_{\mathrm{H}}}{\mathrm{cm}^-3}c\right)$~erg~MeV$^{-2}$~cm$^{-2}$~s$^{-1}$, $\alpha_l~=~2.15415~\pm~0.11482$~. To derive $a_{\mathrm{p}}$ from $a_\gamma$, see \cite{schuppan2012}.}
	\label{fig:fit_G349_11pt}
\end{figure}

To calculate the ionization rate of the molecular cloud, the primary CR proton spectrum at the source is required. To obtain this, observational data about the gamma radiation from instruments as e.g.\ FermiLAT can be used. To generate the gamma radiation observed, the primary CR proton spectrum at the source can be fitted using a proton-proton interaction model like the one from Kelner et al.\ 2006 \cite{kelner_pp2006} or Kamae 2006 \cite{kamae2006, karlsson2008}. As input for these models, typically a (broken) power-law in energy with an exponential cutoff is assumed for the primary CR proton spectrum, which forms a spectrum of gamma rays from the decay of neutral pions formed in interactions of the CR protons with ambient matter. An example of the shape of the primary CR proton spectrum is:
\begin{equation}
	j_{\rm p}(E_{\rm p}) = a_{\rm p}\left(\frac{E_{\rm p}}{E_0}\right)^{-\alpha_l}\left(1+\frac{E_{\rm p}}{E_{\rm br}}\right)^{-(\alpha_h - \alpha_l)} \exp{\left(-\frac{E_{\rm p}}{E_{\rm cutoff}}\right)}~.
\end{equation}
The fit is performed in such a way that the resulting gamma ray spectrum matches the observational gamma data. Free parameters of the fit are the spectral indices below and above the break energy, $\alpha_l$ and $\alpha_h$, the break energy $E_{\mathrm{br}}$, the cut-off energy $E_{\mathrm{cut-off}}$ and the normalization of the resulting gamma ray spectrum, $a_{\gamma}$. 
A fitted gamma ray spectrum for a supernova remnant associated with a molecular cloud is shown in figure \ref{fig:fit_G349_11pt}. 
Since the threshold energy required for the generation of a single pion is about 130~MeV/$c^2$ and a primary CR proton of $E_{\mathrm p}~=~1$~GeV typically gives $\sim~10\%$ of its energy to a secondary produced, the primary CR proton spectrum obtained from modelling the GeV-TeV gamma radiation can only be considered known down to energies of about 1~GeV. The ionization cross section of molecular hydrogen, on the other hand, has its peak at 100~keV and declines rapidly towards higher energies \cite{padovani2009}, so deriving an ionization rate from the modeled primary CR proton spectrum is not trivial. For $E_{\mathrm p}~<~1$~GeV, the spectrum needs to be extrapolated to lower energies carefully, because there is no observational data about this part of the spectrum. In a conservative approach, the primary CR proton spectrum is expected to change from $E_{\mathrm p}^{-\alpha}$ to $E_{\mathrm p}^{-\alpha + 3}$ below 1~GeV. Note that the spectral behavior below $E_{\mathrm p}~<~1$~GeV has no significant influence on the resulting gamma ray spectrum due to the threshold energy for the production of pions.
This modifies the spectral shape of the primary CR proton spectrum as:
\begin{equation}
	j_{\rm p}(E_{\rm p}) = \left\{\begin{array}{cl}
  a_{\rm p}\left(\frac{E_{\rm lb}}{E_0}\right)^{-\alpha_l}\left(1+\frac{E_{\rm lb}}{E_0}\right)^{-(\alpha_h - \alpha_l)} \exp{\left(-\frac{E_{\rm lb}}{E_{\rm cutoff}}\right)} \left(\frac{E_{\rm p}}{E_{\rm lb}}\right)^{s} & (E_{\rm p} \leq E_{\rm lb}) \\
  a_{\rm p}\left(\frac{E_{\rm p}}{E_0}\right)^{-\alpha_l}\left(1+\frac{E_{\rm p}}{E_{\rm br}}\right)^{-(\alpha_h - \alpha_l)} \exp{\left(-\frac{E_{\rm p}}{E_{\rm cutoff}}\right)} & (E_{\rm p} > E_{\rm lb}),\end{array}\right.
  \label{p-spectrum}
\end{equation}
where $s$ is varied from $+1.0$ to $+2.0$~, which should offer a lower limit on the spectrum and therefore on the ionization rate. This is due to the fact that an injection spectrum of $\propto~p^{-\alpha}$, for a loss term of $\dot{p}~\propto~p^{-2}$ like ionization losses, would be modified as $\propto~p^{3-\alpha}$, and the power law indices for the sources discussed are $s~=~1.5~-~2.45$~. \\
The normalization of the primary CR proton spectrum, $a_{\rm p}$, is not trivial to calculate from the resulting gamma ray spectrum normalization $a_{\gamma}$ obtained from fitting. In fact, it requires an estimate of the volume of the source, whereas the gamma ray spectrum normalization does not depend on this quantity. It enters either in the cosmic ray density or in the hydrogen density of the cloud. For a more detailed description of the calculation of the primary CR proton spectrum, see Schuppan et al.\ 2012 \cite{schuppan2012}. 

\section{Predicting the ionization rate}
\label{sec:3}
Using this spectrum, we can calculate the ionization rate due to primary proton ionization following Padovani et al.\ 2009 \cite{padovani2009}. Since the spectral behavior of the primary CR proton spectrum is not precisely known, the ionization rate is calculated for different lower spectral indices $s$ and lower break energies $E_{\mathrm{lb}}$. The predicted ionization rate only weakly depends on the actual spectral index below the lower break, $s$, but rather strongly on the lower break energy $E_{\mathrm{lb}}$. This is due to the strong decline of the ionization cross section of molecular hydrogen for incident proton energies above 100~keV \cite{padovani2009}. 
In our study, we predicted the ionization rates for nine molecular clouds known to be associated with supernova remnants. W49B, in particular, is expected to show signatures of a largely enhanced ionization rate. For at least two objects, the ionization rates are expected to be at least an order of magnitude above the average of Galactic molecular clouds, $\zeta_{\mathrm{gal.~aver}}^{\mathrm{H_2}}~=~2\times10^{-16}$~s$^{-1}$ \cite{Neufeld2010_OHp}. 
In this calculation, there are two quantities left as free parameters. The total energy budget in CR protons interacting with the molecular gas, $W_{\mathrm{p}}$, and the hydrogen density of the cloud, $n_{\mathrm{H}}$. The product of these quantities is fixed from fitting the resulting gamma ray flux to observational data, $W_{\mathrm{p}}n_{\mathrm{H}}~=~const.$, but the values of the quantities themselves cannot be derived. Note that this equation can be rearranged in such a way that $W_{\mathrm{p}}n_{\mathrm{H}}~=~\rho_{\mathrm{p}}V_{\mathrm{cloud}}n_{\mathrm{H}}~=~\rho_{\mathrm{p}}M_{cloud}~=~const.$, where $\rho_{\mathrm{p}}$ is the energy density of the cosmic ray protons. Therefore, the calculation can be adapted to the quantities for which there are the best estimates available. In our calculations, we chose the fixed value of $n_{\mathrm{H}}~=~100$~cm$^{-3}$ as standard value for the clouds. Should future measurements hint at different densities, the results change as $\zeta^{\mathrm{H_2}}~\propto~n_{\mathrm{H}}^{-1}$.

\begin{figure}
	\centering
		\includegraphics[scale=0.8]{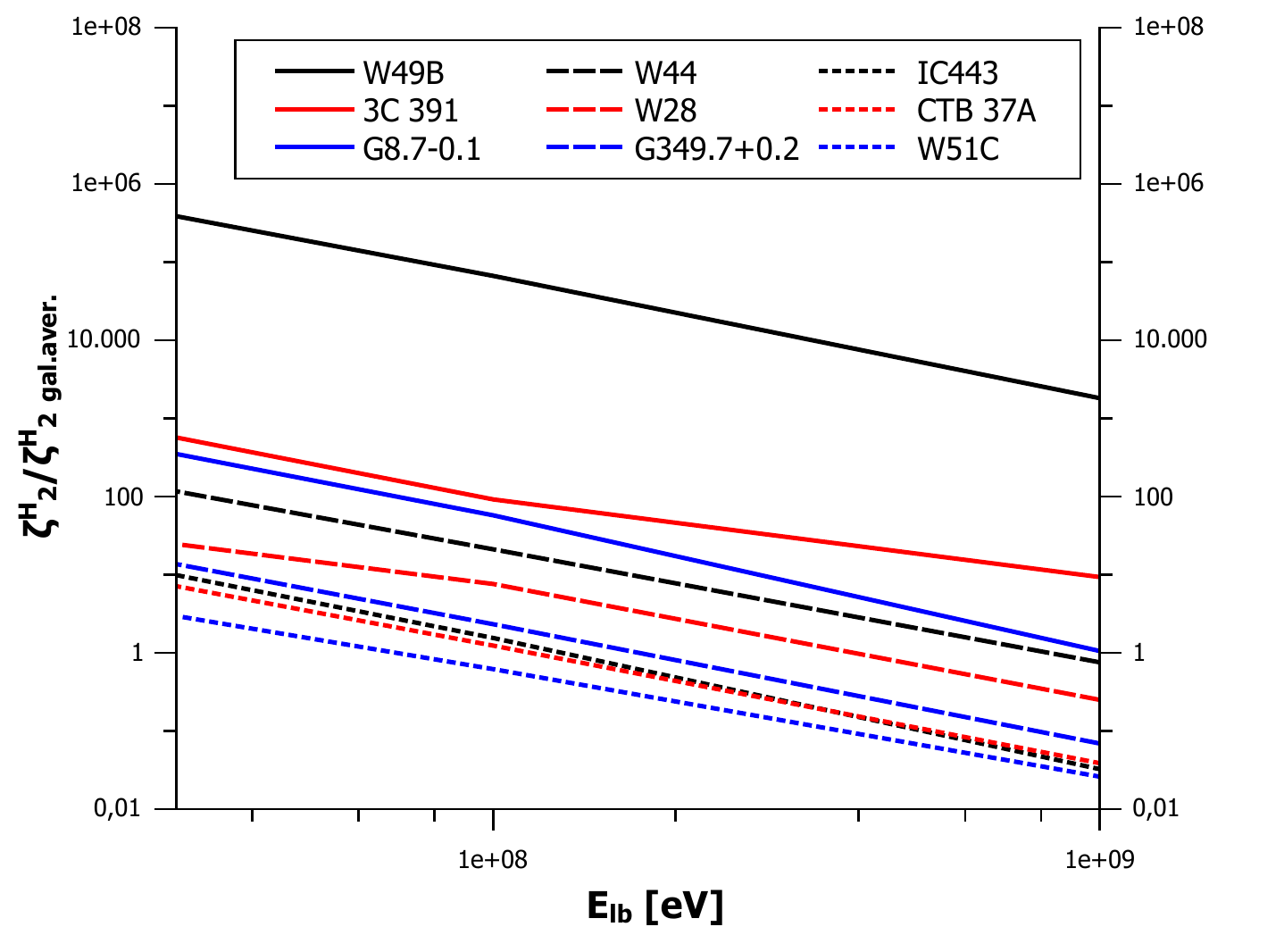}
	\caption{Predicted ionization rate for all objects considered for different lower break energies $E_{\mathrm{lb}}$ of the pirmary CR protons below $E_{\mathrm p}~<~1$~GeV.}
	\label{fig:ion_all}
\end{figure}

This work focuses on the ionization due to primary CR protons, neglecting the contribution of primary CR electrons. For a description of the latter, see Padovani et al.\ 2009 \cite{padovani2009}. Furthermore, the contribution of secondary electrons from primary ionization events is neglected here, which typically enhances the primary ionization rate by a factor of $1.4$ to $2$ \cite{glassgold1973}. Neglecting these processes, the calculation of the ionization rate is performed following Padovani et al.\ 2009 \cite{padovani2009} as:
\begin{equation}
  \zeta^{\rm{H_2}} = \int_{E_{\min}}^{E_{\max}} j_{\rm p}(E_{\rm p})\sigma_{\rm p}^{\rm ion}(E_{\rm p}){\mathrm{d}}E_{\rm p}~, \label{eq_ion}
\end{equation}
where $E_{\min}$ is the minimum energy of protons considered to contribute to the ionization rate and $\sigma_{\rm p}^{\rm ion}$ is the ionization cross section of molecular hydrogen for protons. $E_{\max}$ can be chosen as 1~GeV, increasing it further does not change the resulting ionization rate significantly, due to the rapidly declining ionization cross section for $E_{\mathrm{p}}~>~100$~keV. The value of $E_{\min}$, however, is of large importance for the result. Generally, protons of kinetic energy $E_{\mathrm{p}}~<~100$~keV are not energetic enough to penetrate a cloud of $n_{\mathrm{H}}~=~100$~cm$^{-3}$. Due to the rapid decline of the primary CR proton spectra used here towards energies $E_{\mathrm{p}}~<~E_{\mathrm{lb}}$, a value of $E_{\min}~=~10$~MeV is chosen as a conservative approximation \cite{indriolo2009}. This lower integration limit becomes even more important the lower the value of the lower break energy $E_{\mathrm{lb}}$ is. The latter is varied from $30~$MeV to $1~$GeV in our calculations. Summed up, the larger the lower break energy $E_{\mathrm{lb}}$ and the spectral index $s$ below this lower break energy, the lower the resulting ionization rate. 
In table \ref{parameters}, the parameters used to calculate the ionization rate for all objects considered are shown. The lower break energy, $E_{\mathrm{lb}}$, was altered, while the spectral index of the primary CR proton spectrum below the lower break energy was chosen as $s~=+2.0$ as the most conservative approach. The resulting ionization rate for each object considered depending on the lower break energy is shown in figure \ref{fig:ion_all}. Changing $s$ from $+2.0$ to $+1.0$ does change the results by a factor of $\sim~1.2$ for $E_{\mathrm{lb}}~=~30$~MeV up to a factor of $\sim~2$ for $E_{\mathrm{lb}}~=~1$~GeV. This, in combination with the value of the lower break energy, is likely to be the largest uncertainty in the calculation. As can be seen, for $E_{\mathrm{lb}}~=~100$~MeV, for three objects the predicted ionization rate is at least an order of magnitude above Galactic average for molecular clouds. This choice of the lower break energy $E_{\mathrm{lb}}$ might be the most reasonable. However, even for the most conservative scenario, i.e.\ $E_{\mathrm{lb}}~=~1$~GeV and $s~=~+2.0$, the predicted ionization rate for W49B is orders of magnitude above Galactic average for molecular clouds.

%\begin{landscape}
\begin{table}
	\centering
	{\tiny
		\begin{tabular}[htbp]{|c|c|c|c|c|c|c|c|}
			\hline
  		object & d (kpc) & $V_{\rm cloud}$ (cm$^{3}$) & $E_0$ & $\alpha_l$ & $\alpha_h$ or $E_{\rm cutoff}$ & $W_{\rm p}$ (erg), in protons of $E_{\rm p}$ $>$ 10 MeV & spectral break energy $E_{\rm br}$ (GeV) \\
 		  \hline\hline
  		W51C & 6.0 & $3.3 \times 10^{60}$ & 1 GeV & 1.5 & 2.9 & $7.7 \times 10^{49}$ & 15 \\
  		\hline
  		W44 & 3.0 & $4.2 \times 10^{59}$ & 1 GeV & 1.74 & 3.7 & $1.2 \times 10^{50}$ & 9 \\
  		\hline
  		W28 & 2.0 & $3.2 \times 10^{59}$ & 1 GeV & 1.7 & 2.7 & $3.3 \times 10^{49}$ & 2 \\
  		\hline
  		IC443 & 1.5 & $4.2 \times 10^{59}$ & 1 GeV & 2.0 & 2.87 & $2.4 \times 10^{49}$ & 69 \\
  		\hline
  		W49B & 8.0 & $6.3 \times 10^{56}$ & 1 GeV & 2.0 & 2.7 & $4.4 \times 10^{50}$ & 4 \\
  		\hline
  		G349.7+0.2 & 22.0 & $2.4 \times 10^{59}$ & 1 TeV & 1.7 & 0.16 TeV & $2.2 \times 10^{50}$ & - \\
  		\hline
  		CTB 37A & 11.3 & $3.3 \times 10^{60}$ & 1 TeV & 1.7 & 0.08 TeV & $1.3 \times 10^{50}$ & - \\
  		\hline
  		3C 391 & 8.0 & $3.4 \times 10^{59}$ & 1 TeV & 2.4 & 100 TeV & $2.3 \times 10^{50}$ & - \\
  		\hline
  		G8.7-0.1 & 4.5 & $9.3 \times 10^{59}$ & 1 TeV & 2.45 & 100 TeV & $3.7 \times 10^{50}$ & - \\
  		\hline
		\end{tabular}
		}
\begin{center}		
	\caption{Table of parameters used to calculate the ionization rate. All values for $W_{\rm p}$ are calculated for $n_{\rm{H}}~=~100$~cm$^{-3}$. Gamma ray measurements from: \cite{abdo(W51C)2009, abdo(W44)2010, abdo(W28)2010, abdo(IC443)2010, abdo(W49B)2010, castro2010}\label{parameters}}
\end{center}	
\end{table}
%\end{landscape} 

% Always give a unique label
% and use \ref{<label>} for cross-references
% and \cite{<label>} for bibliographic references
% use \sectionmark{}
% to alter or adjust the section heading in the running head

%Instead of simply listing headings of different levels we recommend to
%let every heading be followed by at least a short passage of text.
%Further on please use the \LaTeX\ automatism for all your
%cross-references and citations.

%Please note that the first line of text that follows a heading is not indented, whereas the first lines of all subsequent paragraphs are.

%Use the standard \verb|equation| environment to typeset your equations, e.g.
%
%\begin{equation}
%a \times b = c\;,
%\end{equation}
%
%however, for multiline equations we recommend to use the \verb|eqnarray| environment\footnote{In physics texts please activate the class option \texttt{vecphys} to depict your vectors in \textbf{\itshape boldface-italic} type - as is customary for a wide range of physical subjects}.
%\begin{eqnarray}
%a \times b = c \nonumber\\
%\vec{a} \cdot \vec{b}=\vec{c}
%\label{eq:01}
%\end{eqnarray}

\section{Ionization signatures}
\label{sec:4}

An enhanced ionization rate of molecular and atomic hydrogen triggers a chemical network, forming about 15 to 20 relatively simple molecular ions. In fully molecular regions, almost every ionization of molecular hydrogen results in formation of an H$_3^+$ molecular ion. Therefore, the detection of this molecular ion is a very good tracer for the ionization rate in a molecular cloud \cite{indriolo2010}. With the ionization rate and a few estimates of cloud parameters like the temperature and density, the amounts of molecular ions formed in a certain ro-vibrational state can be derived using a chemical network and a radiative transfer code as e.g.\ RADEX \cite{Radex2007}. In Becker et al.\ 2011 \cite{becker2011}, the first predicted spectra of H$_2^+$, H$_3^+$, and HeH$^+$ are shown for a reference model cloud. Similar spectra can be predicted for other molecular ions. In general, the most reactive ions, such as H$_2^+$, OH$^+$, and H$_2$O$^+$, will be the most direct tracers of high ionization. Such a prediction can be made for each molecular cloud associated with a supernova remnant. The detection of such relaxation lines in spatial correlation with GeV to TeV gamma radiation is a strong hint at hadronic origin of both signals. The largest uncertainty in the calculation is the shape of the primary CR proton spectrum for energies $E_{\mathrm{p}}~<~1$~GeV. The estimate chosen here is rather conservative, so the derived ionization rates are actually lower limits. A more precise method to predict ionization rates from CR protons is to propagate the CR protons into the cloud, which will be done in future work. Of particular interest is the prediction of ionization profiles, since the profiles induced by CR ionization are expected to be distinct from those induced by photo-ionization and therefore pinpointing the origin of enhanced ionization, at least near the cloud cores, to cosmic rays. With instruments as the Cherenkov Telescope Array and The Atacama Large Millimeter/submillimeter Array, the spatial resolution to perform the correlation study would be available.

\begin{acknowledgement}
We would like to thank R.\ Schlickeiser and P.L.\ Biermann for helpful and inspiring discussions. We also want to thank Marco Padovani for very helpful comments. JBT, SC and FS acknowledge funding from the DFG, Forschergruppe "Instabilities,
Turbulence and Transport in Cosmic Magnetic Fields" (FOR1048, Project BE 3714/5-1) and from the
Junges Kolleg (Nordrheinwestf\"alische Akademie der Wissenschaften und der K\"unste). JHB is grateful to the Swedish National Space Board for
support. We further acknowledge the support by the
Research Department of Plasmas with Complex Interactions (Bochum).
\end{acknowledgement}


\begin{thebibliography}{99.}%
\bibitem{abdo(W51C)2009} {Abdo}, A.~A. and others: "Fermi LAT Discovery of Extended Gamma-Ray Emission in the Direction of Supernova Remnant W51C", Astroph.~J.~Lett. \textbf{706}, L1 (2009)

\bibitem{abdo(W44)2010} {Abdo}, A.~A. and others: "Gamma-Ray Emission from the Shell of Supernova Remnant W44 Revealed by the Fermi LAT", Science \textbf{327}, 1103 (2010)

\bibitem{abdo(W28)2010} {Abdo}, A.~A. and others: "Fermi Large Area Telescope observations of the supernova remnant W28 (G6.4-0.1)", Astroph.~J.~Lett. \textbf{718}, 348 (2010)

\bibitem{abdo(IC443)2010} {Abdo}, A.~A. and others: "Observation of supernova remnant IC 443 with the Fermi Large Area Telescope", Astroph.~J.~Lett. \textbf{712}, 459 (2010)

\bibitem{abdo(W49B)2010} {Abdo}, A.~A. and others: "Fermi-LAT Study of Gamma-ray Emission in the Direction of Supernova Remnant W49B", Astroph.~Journal \textbf{722}, 1303 (2010)

\bibitem{becker2011} {Becker},  J.~K. et~al.: "Tracing the sources of cosmic rays with molecular ions",  Astroph.~J.~Lett. \textbf{739}, L43 (2011)

\bibitem{black2007} {{Black}, J.~H.}: "The short but exciting lives of molecular ions in interstellar clouds". In: J.~L.~Lemaire~and~F.~Combes (eds.),  Molecules in Space and Laboratory, (2007)

\bibitem{castro2010} {{Castro}, D. and {Slane}, P.}: "Fermi Large Area Telescope Observations of Supernova Remnants Interacting with Molecular Clouds", Astroph.~Journal \textbf{717}, 372 (2010)

\bibitem{glassgold1973} {{Glassgold}, A.~E. and {Langer}, W.~D.}: "Heating of Molecular-Hydrogen Clouds by Cosmic Rays and X-Rays", Astroph.~Journal \textbf{186}, 859 (1973)

\bibitem{indriolo2009} {{Indriolo}, N. and others}: "The Implications of a High Cosmic-Ray Ionization Rate in Diffuse Interstellar Clouds", Astroph.~Journal \textbf{694}, 257 (2009)

\bibitem{indriolo2010} {{Indriolo}, N. and others}: "Investigating the Cosmic-ray Ionization Rate Near the Supernova Remnant IC 443 through H$^{+}$ $_{3}$ Observations", Astroph.~Journal \textbf{724}, 1357 (2010)

\bibitem{karlsson2008} {{Karlsson}, N.\ and {Kamae}, T.}: "Parameterization of the Angular Distribution of Gamma Rays Produced by p-p Interaction in Astronomical Environments", Astroph.~Journal \textbf{674}, 278 (2008)

\bibitem{kamae2006} {Kamae}, T. and others: "Parameterization of $\gamma$, e$^{+/-}$, and Neutrino Spectra Produced by p-p Interaction in Astronomical Environments", Astroph.~Journal \textbf{647}, 692 (2006)

\bibitem{kelner_pp2006} {Kelner}, S.~R. and others: "Energy spectra of gamma rays, electrons, and neutrinos produced at proton-proton interactions in the very high energy regime", Phys.~Rev.~D. \textbf{74:3} (2006)

\bibitem{mccarthy2006} {{McCarthy}, M.~C. and others}:  "Laboratory and Astronomical Identification of the Negative Molecular Ion C$_{6}$H$^{-}$",  Astroph.~J.~Lett. \textbf{652}, L141 (2006)

\bibitem{Neufeld2010_OHp} {Neufeld}, D.~A.\ et~al.: "Herschel/HIFI observations of interstellar OH$^{+}$ and H$_{2}$O$^{+}$ towards W49N: a probe of diffuse clouds with a small molecular fraction", Astron.~\&~Astroph. \textbf{521}, L10 (2010)

\bibitem{padovani2009} {Padovani}, M. and others: "Cosmic-ray ionization of molecular clouds",  Astron.~\&~Astroph. \textbf{501}, 619 (2009)

\bibitem{schuppan2012} {Schuppan}, F. and others: "Cosmic-ray-induced ionization in molecular clouds adjacent to supernova remnants - Tracing the hadronic origin of GeV gamma radiation", Astron.~\&~Astroph. \textbf{541} A126 (2012)

\bibitem{Radex2007} {van der Tak}, F.~F.~S. and others.: "A computer program for fast non-LTE analysis of interstellar line spectra. With diagnostic plots to interpret observed line intensity ratios", Astron.~\&~Astroph. \textbf{468}, 627 (2007)

% Use the following syntax and markup for your references if 
% the subject of your book is from the field 
% "Mathematics, Physics, Statistics, Computer Science"
%
% Contribution 
%\bibitem{science-contrib} Broy, M.: Software engineering --- from auxiliary to key technologies. In: Broy, M., Dener, E. (eds.) Software Pioneers, pp. 10-13. Springer, Heidelberg (2002)
%
% Online Document
%\bibitem{science-online} Dod, J.: Effective substances. In: The Dictionary of Substances and Their Effects. Royal Society of %Chemistry (1999) Available via DIALOG. \\
%\url{http://www.rsc.org/dose/title of subordinate document. Cited 15 Jan 1999}
%
% Monograph
%\bibitem{science-mono} Geddes, K.O., Czapor, S.R., Labahn, G.: Algorithms for Computer Algebra. Kluwer, Boston (1992) 
%
% Journal article
%\bibitem{science-journal} Hamburger, C.: Quasimonotonicity, regularity and duality for nonlinear systems of partial differential %equations. Ann. Mat. Pura. Appl. \textbf{169}, 321--354 (1995)
%
% Journal article by DOI
%\bibitem{science-DOI} Slifka, M.K., Whitton, J.L.: Clinical implications of dysregulated cytokine production. J. Mol. Med. (2000) doi: %10.1007/s001090000086 
%
%\bigskip

% Use the following (APS) syntax and markup for your references if 
% the subject of your book is from the field 
% "Mathematics, Physics, Statistics, Computer Science"
%
% Online Document
%\bibitem{phys-online} J. Dod, in \textit{The Dictionary of Substances and Their Effects}, Royal Society of Chemistry. (Available via %DIALOG, 1999), 
%\url{http://www.rsc.org/dose/title of subordinate document. Cited 15 Jan 1999}
%
% Monograph
%\bibitem{phys-mono} H. Ibach, H. L\"uth, \textit{Solid-State Physics}, 2nd edn. (Springer, New York, 1996), pp. 45-56 
%
% Journal article
%\bibitem{phys-journal} S. Preuss, A. Demchuk Jr., M. Stuke, Appl. Phys. A \textbf{61}
%
% Journal article by DOI
%\bibitem{phys-DOI} M.K. Slifka, J.L. Whitton, J. Mol. Med., doi: 10.1007/s001090000086
%
% Contribution 
%\bibitem{phys-contrib} S.E. Smith, in \textit{Neuromuscular Junction}, ed. by E. Zaimis. Handbook of Experimental Pharmacology, %vol 42 (Springer, Heidelberg, 1976), p. 593
%
%\bigskip
%

\end{thebibliography}
\end{document}